\documentclass[amsmath,amssymb,pre,twocolumn,superscriptaddress,longbibliography,hypertext]{revtex4-2}

\usepackage{float}
\usepackage{siunitx}
\usepackage{booktabs}
\usepackage{graphicx}
\usepackage{array}
\usepackage{amsthm}
\usepackage{enumerate}
\usepackage[table]{xcolor}
\usepackage{bbm}
\usepackage{color}
\usepackage[normalem]{ulem}
\usepackage{multirow}
\usepackage{soul}
\RequirePackage[colorlinks=true,linkcolor=blue]{hyperref}
\usepackage{braket}

\begin{document}

\title{Geometry and dynamical morphology of growing bacterial colonies}
\author{Benjamin Evert Himberg}
\affiliation{Department of Physics, University of  Vermont, Burlington, VT 05405}
\affiliation{Materials Science Program, University of  Vermont, Burlington, VT 05405}
\email{bhimberg@uvm.edu}
\author{Sanghita Sengupta}
\affiliation{Department of Chemistry, Brandeis University, Waltham, MA 02453}
\email{ssengupta@brandeis.edu}

\date{\today}        

\begin{abstract}
We study non-equilibrium bacterial colony growth using a geometry-first, time-resolved analysis of morphology. From time-lapse microscopy data, we track the coupled evolution of area, perimeter, and boundary-sensitive shape descriptors along the full growth history. We find that non-equilibrium growth can exhibit extended intervals of compact area–perimeter scaling with exponent $\alpha\approx2$, consistent with growth governed by a single effective geometric length scale, as well as time-localized breakdowns of this scaling during ongoing growth. These breakdowns coincide with transient boundary reorganization while bulk area growth remains sustained. Our results demonstrate that visually distinct morphologies can arise within the same geometric growth regime, and that departures from single-scale behavior reflect intrinsic dynamical restructuring rather than growth arrest. More broadly, this work establishes time-resolved geometry as a coarse-grained framework for identifying when non-equilibrium growth departs from single-scale geometric constraints in living systems.
\end{abstract}

\maketitle

\section{Introduction}
\label{sec:intro}

Patterns are ubiquitous in natural systems, spanning length and time scales from crystalline snowflakes and sand dunes to collective motion in flocks and spatial organization in biological populations \cite{turing1952chemical, ball1998self, ball2012pattern, langer1980instabilities, Schweisguth2019}. In many non-living systems, the emergence of structure can be understood within the framework of free-energy minimization or equilibrium statistical mechanics \cite{landau1980statistical}. Living systems, by contrast, operate far from equilibrium: growth and organization are sustained by continuous driving and dissipation, and the resulting patterns reflect irreversible dynamics rather than relaxation toward equilibrium states \cite{ramaswamy2010mechanics, marchetti2013hydrodynamics, cross1993pattern}.

Bacterial colonies provide a particularly accessible and experimentally tractable example of non-equilibrium pattern formation \cite{Ben-Jacob01091993, Ben-Jacob01051997, ben-jacob95}. As colonies grow, they exhibit diverse morphologies, boundary structures, and internal organization across strains and conditions \cite{Golding1998, Matsushita1990}. From a physical perspective, colony growth shares features with a broad class of non-equilibrium growth processes, including kinetic roughening, aggregation, and driven interface dynamics \cite{KPZ1986, Barabasi1995, Krug1997}. Unlike idealized growth models, however, living colonies continuously reorganize both their bulk and boundary structure during growth, producing morphologies that are not fixed by a single compact or ramified growth mode \cite{BenJacob2006, Farrell2013}. Capturing this evolving structure in a quantitative and physically interpretable manner remains a central challenge \cite{pokhrel, Meakin1998}.

A substantial body of work has approached bacterial colony morphology through mechanistic or model-based descriptions of growth, incorporating factors such as nutrient transport, cellular interactions, or specific growth rules \cite{Tronnolone2018, fujikawa1989fractal, muller2002, calvo, grimson, kozlovsky, ben-jacob95}. While these approaches have yielded valuable insights, a complementary question remains largely open: to what extent can colony morphology be characterized, compared, and distinguished using geometry alone, independent of microscopic growth mechanisms? In particular, it is not obvious whether visually distinct colony shapes necessarily reflect fundamentally different non-equilibrium growth dynamics, or whether they can arise within a shared geometric framework subject to controlled departures.

Most existing studies of colony morphology rely on static descriptors of mature structures, such as final shape, roughness, or fractal dimension \cite{fujikawa1989fractal, ben-jacob95, Meakin1998}. While such measures provide useful summaries, they obscure the dynamical pathways through which morphology develops and cannot distinguish whether similar final morphologies arise from distinct growth histories or identify transient reorganizations during growth. A framework that treats morphology itself as a time-dependent observable is therefore required.

In non-equilibrium statistical physics, geometric scaling relations such as area--perimeter scaling and fractal measures have long been used to characterize growth processes ranging from Eden growth to diffusion-limited aggregation \cite{Eden1961, WittenSander1981, Meakin1998}. In biological contexts, these tools are typically applied as static descriptors \cite{fujikawa1989fractal, ben-jacob95, Golding1998}. Treating them dynamically, however, offers a route to identifying when growth is governed by a single effective geometric length scale and when additional geometric degrees of freedom emerge \cite{Krug1997, cross1993pattern}. In particular, compact growth controlled by a single length scale implies a fixed geometric relation between bulk accumulation and boundary advance, while deviations from this relation signal geometric reorganization during growth. Importantly, different geometric observables encode complementary and non-redundant information: area reflects bulk accumulation, perimeter probes boundary activity, fractal dimension captures multiscale roughness, and circularity quantifies global symmetry and compactness \cite{Mandelbrot1982}.

The geometric analysis presented here is implemented using \texttt{PyPetana}\cite{sanghita_pypetana_2024}, an open-source Python framework built on OpenCV\cite{bradski2008learning} for quantitative analysis of growth morphology from time-resolved images. Related geometry-based approaches have recently been applied to other living systems, such as \textit{Physarum}, to characterize morphological growth using descriptors including area, perimeter, circularity, and fractal dimension \cite{bajpai2025morphologicalcomputationalcapacityphysarum}. In the present work, we extend this approach to collective bacterial colonies and apply it to capture heterogeneous and dynamically reorganizing growth across long time scales. While this framework enables systematic extraction of geometric observables, the focus of this study is on the physical interpretation of dynamical morphology rather than on computational methodology.

We adopt a geometry-first approach to bacterial colony growth, treating morphology as a dynamical observable rather than a static descriptor. We analyze time-resolved images of five bacterial strains exhibiting visually distinct colony morphologies but comparable growth coverage. By characterizing growth through the coupled evolution of area, perimeter, and dimensionless shape descriptors, we ask whether geometry can reveal both shared constraints and meaningful distinctions in non-equilibrium growth behavior. Unlike prior work that applies geometric measures to final or time-averaged morphologies, our approach resolves when geometric constraints hold and when they fail. In particular, the time-ordered area--perimeter trajectory distinguishes continuous compact growth from transient geometric restructuring even when growth is sustained and area increases monotonically.

Applying this framework, we uncover a striking separation between visual morphology and geometric characterization. Although colony shape and boundary organization differ markedly across strains, the time-ordered area–perimeter trajectories reveal when compact geometric constraints persist and when they break down. In particular, departures from the conventional compact area–perimeter scaling associated with ramified two-dimensional growth coincide with correlated excursions in boundary-based shape descriptors, signaling the activation of additional geometric degrees of freedom during growth, even as colony area increases monotonically.


The remainder of the paper is organized as follows. In Sec.~\ref{sec:methods}, we describe the image-based extraction of geometric observables and the construction of shape descriptors. In Sec.~\ref{sec:results}, we present the time evolution of area, perimeter, effective fractal dimension, and global shape parameters, together with area--perimeter scaling analysis of non-equilibrium growth. We conclude with a summary and outlook in Sec.~\ref{sec:discussion}.

\section{Methods}
\label{sec:methods}

\begin{figure*}
    \centering
    \includegraphics[width=0.95\linewidth]{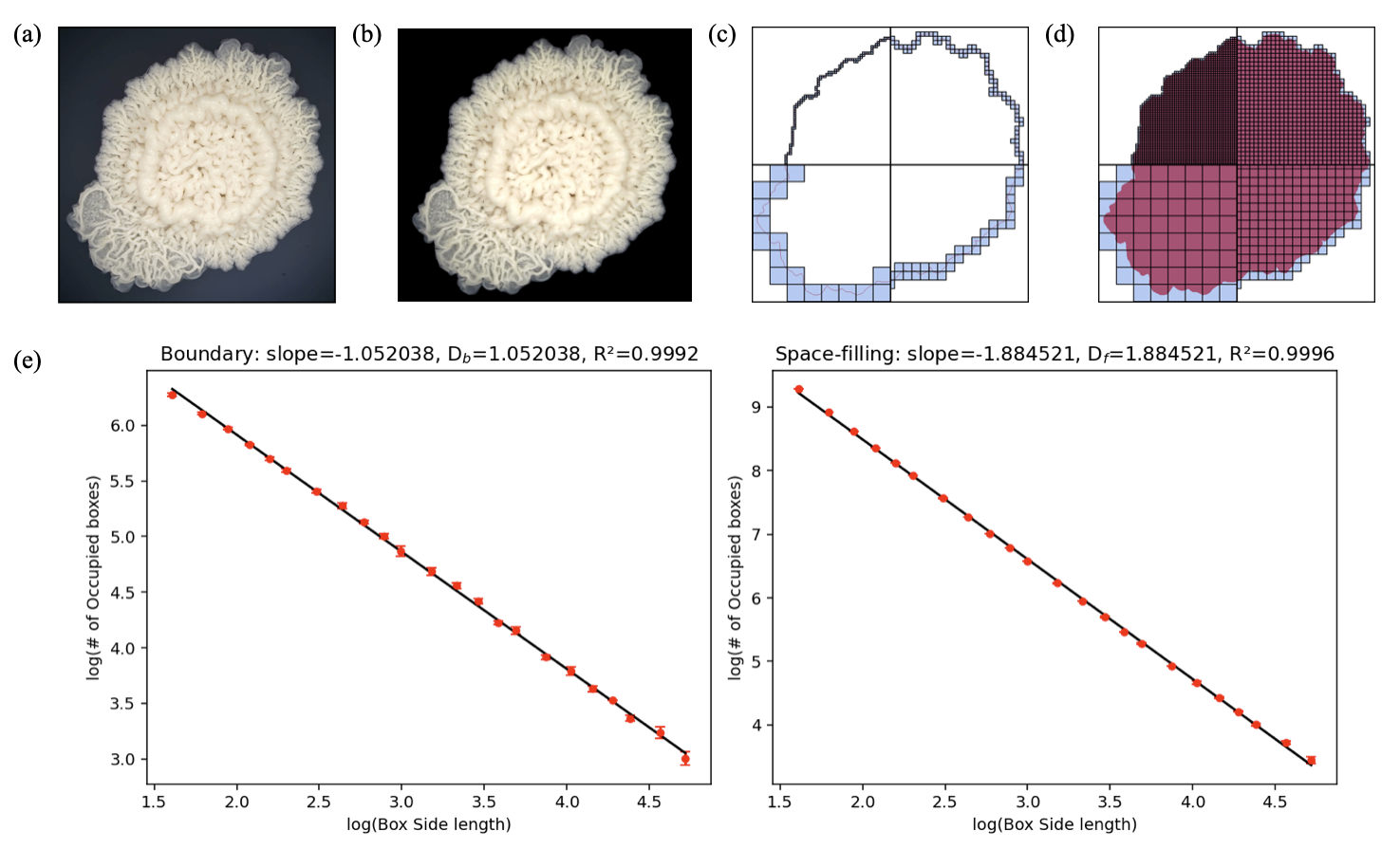}
    \caption{
Geometric analysis pipeline for extracting time-resolved observables from colony images.
(a) Representative microscopy image at a single time point.
(b) Binary mask obtained through preprocessing and adaptive thresholding; the enclosed area $A(t)$ and perimeter $P(t)$ are computed from the segmented footprint and its contour, respectively.
(c) Box-counting construction applied to the extracted colony boundary to estimate the effective boundary fractal dimension $D_b$, using box sizes $\epsilon = 5, 10, 20,$ and $40$ pixels, based on the scaling relation $N(\epsilon)\sim \epsilon^{-D_b}$.
(d) Box-counting applied to the filled colony mask to estimate the space-filling (mass) fractal dimension $D_f$, using the same box sizes.
(e) Log--log box-counting plots for the boundary and filled colony interior, yielding estimates of $D_b$ and $D_f$.
All fractal-dimension analyses are performed using a boundary thickness of one pixel.
}
\label{fig:methods}
\end{figure*}

In this section, we describe the extraction and construction of geometric observables and global shape descriptors used to characterize the non-equilibrium morphological evolution of growing bacterial colonies. Our approach treats colony morphology as a dynamical object defined at the level of geometric observables rather than as a static outcome. All observables are extracted directly from time-resolved images without assuming specific microscopic growth mechanisms. We begin by describing the image-based extraction of geometric observables from evolving colony shapes.

\subsection{Image-based extraction of geometric observables}
\label{subsec:image}

Time-resolved microscopy videos of growing \textit{Bacillus subtilis} colonies for strains 100, 102, 106, 108 and NCIB3610 were obtained from the supplementary materials of Ref.~\cite{Futo2022_SciRep}. The recordings were acquired at regular time intervals of 6 minutes under standardized experimental conditions and span multiple days of continuous growth~\cite{Futo2022_SciRep}.

Image processing and extraction of geometric observables were performed using \texttt{PyPetana}\cite{sanghita_pypetana_2024,sengupta_pypetana_doc_2024}, an open-source Python-based pipeline built on the OpenCV library\cite{bradski2008learning}. Raw experimental videos were segmented into individual frames corresponding to successive stages of colony expansion. Images were converted to grayscale and preprocessed to enhance contrast between the colony and background. Detection of the outer boundary of the colony footprint was performed using adaptive thresholding, followed by morphological operations within the OpenCV library\cite{sanghita_pypetana_2024}.

Colony boundaries were extracted from the resulting binary masks using contour-detection algorithms (see Fig.~\ref{fig:methods} for details). This contour was used consistently for all subsequent geometric analyses. Representative raw images, processed masks, extracted time series, and segmentation parameters for all strains are provided in Ref.~\cite{BacterialColonyGeometryData_Zenodo}.

Across the five strains analyzed here, the image-extraction procedure yielded more than 4000 segmented colony shapes sampled along evolving growth morphologies, with each strain contributing at least several hundred distinct morphological configurations. Importantly, these frames do not represent independent biological realizations, but successive states along a single non-equilibrium growth trajectory of each colony. This dense temporal sampling enables robust characterization of evolving geometry while preserving transient dynamical features of the morphology.

In the next subsection, we introduce the geometric observables extracted from the experimental images and describe how they relate to the shape descriptors used to characterize the evolving colony morphologies.

\subsection{Characterization of colony growth via geometric shape factors in morphology space}
\label{subsec:geometry}
Geometric and morphological observables are extracted from segmented colony images, where each time frame yields a binary mask that provides a consistent geometric representation from which global shape, bulk, and boundary observables are computed without assuming specific microscopic growth mechanisms (Fig.~\ref{fig:methods}).

Corresponding to each mask, we compute the enclosed area $A(t)$ and the perimeter $P(t)$ using standard contour-based methods. Area captures bulk accumulation within the colony morphology, while perimeter probes the organization of the outer growth front at the image resolution. We first characterize global shape symmetry using the circularity,
\begin{equation}
C(t) = \frac{4\pi A(t)}{P^{2}(t)},
\end{equation}
where $C=1$ corresponds to a perfect circle and smaller values indicate increasing deviation from circular symmetry due to large-scale anisotropy and boundary corrugation.

To quantify the geometric cost associated with boundary corrugation relative to bulk filling, we also define the compactness, 
\begin{equation}
\xi(t) = \frac{P^{2}(t)}{A(t)}.
\end{equation}
Although $\xi(t)$ is formally the reciprocal of circularity, the two observables emphasize different geometric aspects of growth. The circularity $C(t)$ is a bounded, normalized measure of global shape symmetry, whereas $\xi(t)$ is an unbounded quantity that accentuates deviations from compact growth by weighting perimeter relative to area. In this sense, $\xi(t)$ serves as a sensitive geometric diagnostic of boundary complexity during colony growth.

We next characterize multiscale boundary organization by computing an effective boundary fractal dimension $D_{b}(t)$ using a box-counting approach \cite{Mandelbrot1982,Meakin1998}, as illustrated in Fig.~\ref{fig:methods}(c,e). For each time frame, the extracted colony boundary is represented as a boundary-only binary mask. To ensure topological connectivity and suppress pixel-scale discretization artifacts, the boundary was dilated to a thickness of one pixels prior to box counting. The number of occupied boxes $N(\epsilon)$ intersecting the boundary is then computed over a range of box sizes $\epsilon$, and the effective boundary dimension is estimated from the scaling relation\cite{Mandelbrot1982}
\begin{equation}
N(\epsilon) \sim \epsilon^{-D_b}.
\end{equation}
Box-counting fits were restricted to box sizes $\epsilon \ge 5$ pixels, thereby excluding both pixel-scale noise at small $\epsilon$ and finite-size saturation at large $\epsilon$ (see Fig.~\ref{fig:methods}(e)). 


In addition to boundary organization, we also quantify the degree of interior space filling by computing a bulk (mass) fractal dimension $D_f(t)$ from the filled colony masks (see Fig.~\ref{fig:methods}(d,e)). For each time frame, the full binary mask representing the occupied colony area is subjected to a box-counting analysis analogous to that used for the boundary. The number of occupied boxes $N_f(\epsilon)$ covering the filled region is measured as a function of box size $\epsilon$, and the bulk fractal dimension is estimated from the scaling relation\cite{Mandelbrot1982}
\begin{equation}
N_f(\epsilon) \sim \epsilon^{-D_f}.
\end{equation}
Unlike the boundary dimension $D_b$, which characterizes multiscale roughness of the interface, $D_f$ probes how the colony interior occupies space across scales and therefore serves as a measure of bulk packing or space-filling efficiency. For compact two-dimensional objects, $D_f \approx 2$, while smaller values indicate increasingly ramified or heterogeneous interior structure. 


We then assess whether colony growth can be characterized by a single effective geometric length scale by examining the scaling relation between area and perimeter over the full growth trajectory. Specifically, we test a scaling form of the type\cite{cael}
\begin{equation}\label{eq:ap}
A \sim P^{\alpha},
\end{equation}
where the exponent $\alpha$ is obtained from a linear fit to $\log A$ versus $\log P$ using data from the entire time-resolved trajectory for each strain. Because the perimeter entering this relation is measured at a fixed image resolution, the extracted exponent $\alpha$ should be interpreted as a single-scale geometric diagnostic, rather than as a multiscale fractal boundary exponent. Accordingly, deviations from a constant $\alpha$ along the trajectory indicate departures from growth governed by a single effective geometric scale.


In the next section, we present results characterizing the temporal evolution of non-equilibrium growth using the geometric observables introduced above.


\section{Results}
\label{sec:results}

\begin{figure*}
    \centering
    \includegraphics[width=0.98\linewidth]{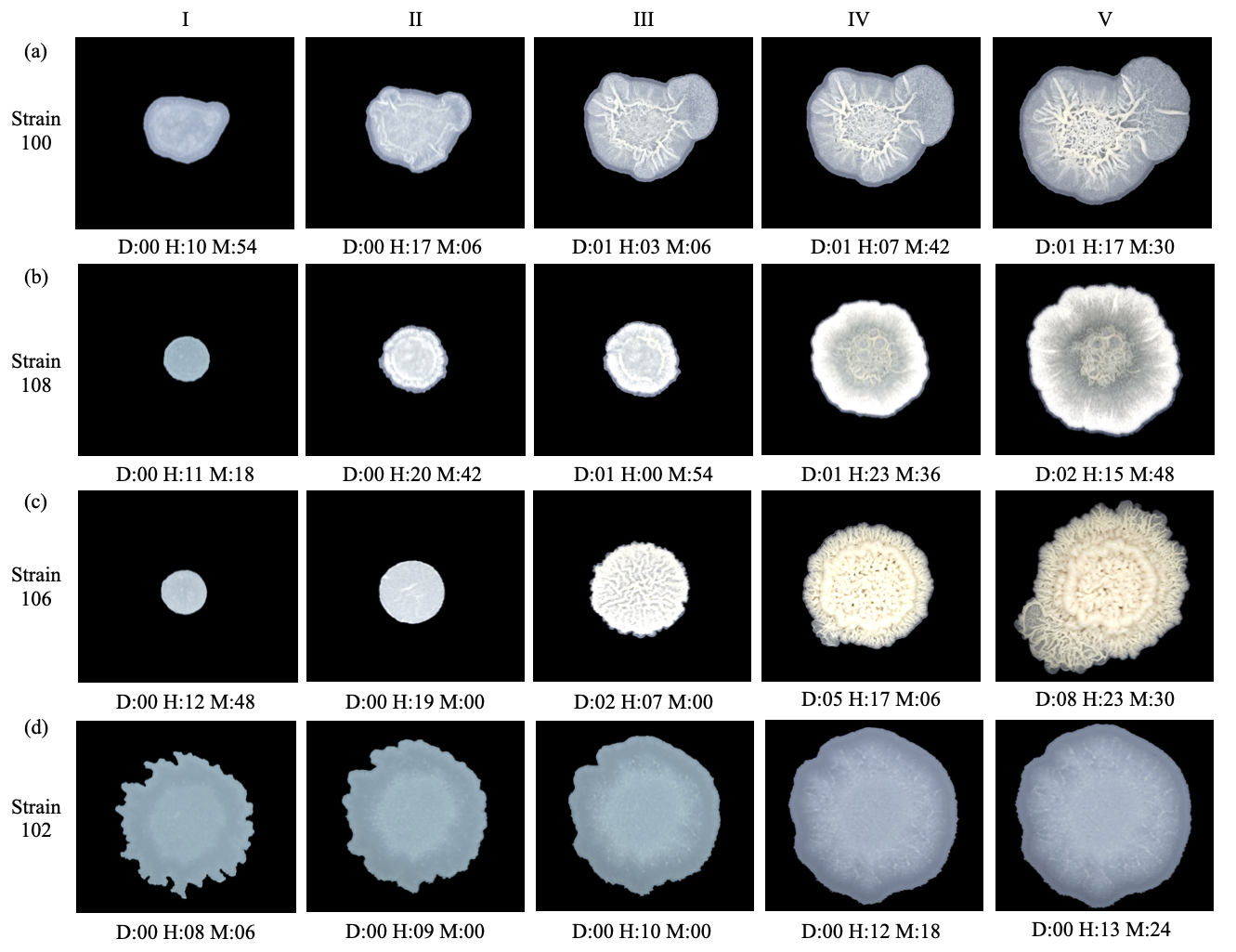}
    \caption{Time-resolved snapshots of colony growth morphology for four \emph{Bacillus subtilis} strains:
(a) strain 100, (b) strain 108, (c) strain 106, and (d) strain 102.
Five representative time points (I)--(V) are shown for each strain, spanning early to late stages of growth.
The colony footprints shown here correspond to those used in the geometric analysis.
These snapshots provide a qualitative reference for the temporal evolution of colony shape, boundary organization, and internal patterns across strains.}
    \label{fig:morphology}
\end{figure*}
Using the methodology described in Sec.~\ref{sec:methods}, we analyze time-resolved geometric observables extracted from experimental images of growing bacterial colonies. We begin with representative growth morphologies for individual strains and their corresponding temporal evolution of area and perimeter. We then examine the time dependence of multiple boundary and shape descriptors, including circularity $C$, compactness $\xi$, mass fractal dimension $D_{f}$, and boundary fractal dimension $D_{b}$. Finally, by constructing time-ordered area–perimeter trajectories, we demonstrate how visually distinct growth morphologies can exhibit identical compact-growth scaling, while time-localized extrema in geometric descriptors coincide with a breakdown of this scaling.

\subsection{Time-resolved qualitative colony growth morphology}
\label{subsec:snapshots}

We begin by examining representative time-resolved snapshots of colony morphology for each strain at selected times spanning early, intermediate, and late stages of growth (Fig.~\ref{fig:morphology}). These snapshots provide a qualitative reference for the strain-dependent diversity of colony shapes and serve to anchor the quantitative geometric analysis presented below.
Figure~\ref{fig:morphology} shows representative colony morphologies for strains 100, 108, 106, and 102 at five distinct time points, progressing from early to late stages of growth (see Appendix~\ref{sec:strain-NC} for strain NCIB3610). Extracted colony boundaries illustrate the contours used for subsequent geometric measurements. While representative snapshots are shown here, the full time-resolved geometric data for all strains are available in the associated Zenodo dataset~\cite{BacterialColonyGeometryData_Zenodo}. For all strains, each dataset contains several hundred time frames. 
Across all strains, early-stage colonies deviate measurably from ideal circular symmetry, as evident from the first column of Fig.~\ref{fig:morphology}. As growth proceeds, strain-dependent heterogeneity emerges in both boundary structure and internal patterning. These qualitative differences persist and evolve over time, resulting in visually distinct late-stage morphologies across strains. Importantly, as shown in the following sections, such visual diversity alone does not uniquely determine the underlying geometric scaling behavior, motivating a time-resolved quantitative analysis of area, perimeter, and boundary descriptors.

\subsection{Growth kinematics: Time evolution of area--perimeter}
\label{subsec:growth_kinematics}

\begin{figure}[htp!]
    \centering
    \includegraphics[width=0.98\columnwidth]{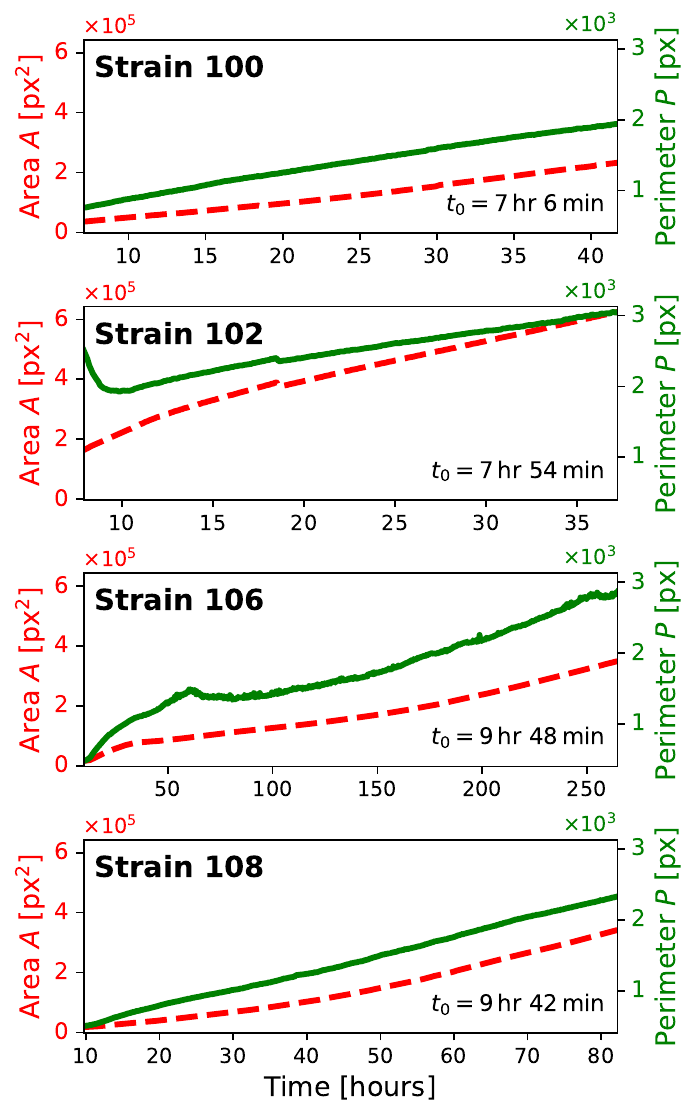}
    \caption{Time evolution of colony area $A(t)$ (red, dashed) and perimeter $P(t)$ (green, solid) for multiple bacterial strains, shown in separate panels. Time is reported in hours, with $t_{0}$ denoting the first time point at which the colony boundary is reliably identified. Area and perimeter are plotted on separate vertical axes to emphasize the concurrent evolution of bulk growth and boundary morphology. For all strains, both quantities increase overall with time, indicating sustained growth without fragmentation or arrest, while strain-dependent temporal features are most pronounced in the perimeter. All strains are analyzed over comparable spatial extents, with each panel spanning the full observation window for the corresponding strain. Non-monotonic features in the perimeter signal transient boundary reorganizations that are not apparent from area growth alone.}
    \label{fig:AP-kinematics}
\end{figure}
We begin by establishing the basic kinematic features of colony growth. Representative snapshots of colony morphology for each strain at selected times are shown in Fig.~\ref{fig:morphology} and Fig.~\ref{fig:morph-nc}, illustrating pronounced qualitative differences in colony shape and internal structure. Despite these visual differences, all colonies exhibit continuous growth without fragmentation or arrest over the full observation window.
To quantify bulk and boundary growth, we examine the time evolution of the colony area $A(t)$ and perimeter $P(t)$ (Fig.~\ref{fig:AP-kinematics}). For all strains, both quantities increase smoothly and monotonically in time, indicating sustained growth. Area growth remains comparatively regular across strains, whereas the perimeter exhibits nontrivial temporal features that reflect changes in boundary morphology. Results for strain NCIB3610 are shown separately in Appendix~\ref{sec:strain-NC}.
Notably, strain 106 exhibits a pronounced, time-localized peak in the perimeter over the interval $t \sim 49$–$100$~h. This feature coincides with the emergence of highly corrugated boundary structures visible in columns (III)–(IV) of Fig.~\ref{fig:morphology}, indicating a transient reorganization of the colony boundary. At later times, the perimeter decreases relative to area growth as the colony evolves toward a distinct morphology.
A similar, though more rapid, boundary reorganization is observed for strain 102 at early times. Here, the perimeter displays an initial peak corresponding to the amoeboid, lobed morphologies shown in columns (I)–(II) of Fig.~\ref{fig:morphology}, followed by a reduction as the boundary visibly smooths while bulk growth continues.
Together, these observations highlight a separation between smooth bulk expansion and dynamically evolving boundary structure. Even when area growth appears regular, the perimeter retains signatures of transient morphological reorganizations. This motivates a geometric analysis that explicitly distinguishes bulk size from boundary complexity, which we pursue in the following sections.

\subsection{Time evolution of shape order parameters}
\label{subsec:morphology_time}
\begin{figure*}[htp!]
    \centering
    \includegraphics[width=0.98\linewidth]{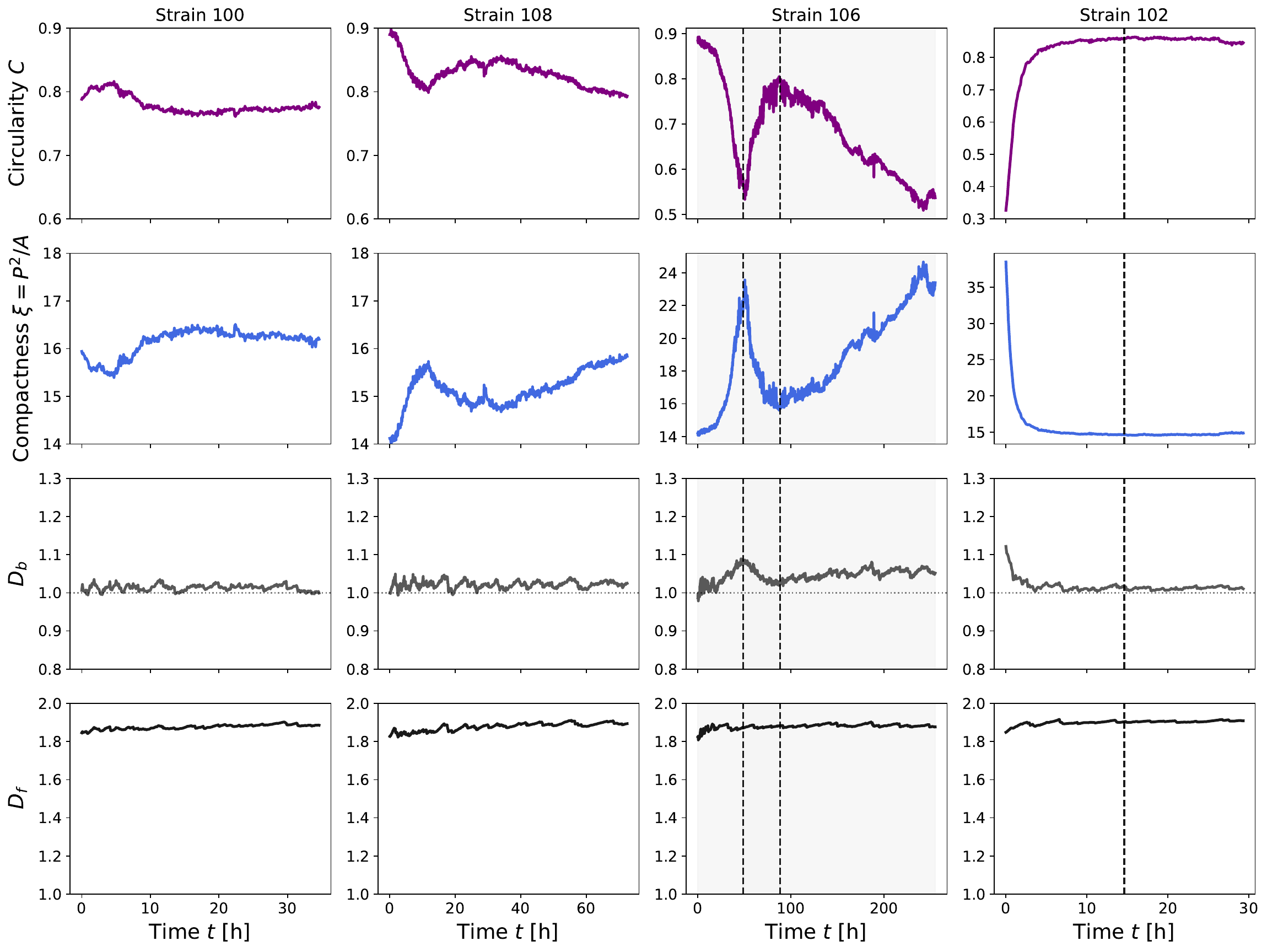}
    \caption{Temporal evolution of shape order parameters for all strains. Shown are the circularity $C(t)$ (top row), compactness $\xi(t)=P^{2}/A(t)$ (second row), effective boundary fractal dimension $D_{b}(t)$ (third row), and mass fractal dimension $D_{f}(t)$ (bottom row) for strains 100, 108, 106, and 102 (columns left to right). Boundary-sensitive descriptors ($C$, $\xi$, $D_{b}$) exhibit strain-dependent temporal structure, while the mass fractal dimension $D_{f}$ evolves more gradually toward values characteristic of compact two-dimensional growth. For strain 106, vertical dashed lines mark characteristic times associated with extrema in $C(t)$, coinciding with correlated responses across boundary-sensitive descriptors and with the morphological configurations shown in Fig.~\ref{fig:morphology}(c). For strain 102, a vertical dashed line indicates the onset of stabilization in boundary-sensitive descriptors following a rapid initial reorganization, beyond which all observables evolve smoothly.}

    \label{fig:shape}
\end{figure*}

To characterize colony morphology beyond global size, we analyze the time evolution of dimensionless shape descriptors: circularity $C(t)$, compactness $\xi(t)$, space-filling fractal dimension $D_{f}(t)$, and boundary fractal dimension $D_{b}(t)$, as defined in Sec.~\ref{subsec:geometry}. These observables probe complementary aspects of colony morphology, ranging from global symmetry to boundary roughness and space-filling properties.

Figure~\ref{fig:shape} shows the temporal evolution of these shape descriptors for all strains. For strains 100 and 108, all observables evolve smoothly with relatively modest fluctuations. Circularity remains within the range $C \sim 0.75$--$0.9$, indicating morphologies that remain close to compact, approximately circular shapes throughout growth. Over the same period, the mass fractal dimension $D_{f}$ increases gradually toward saturation, remaining within $D_{f} \approx 1.6$--$1.9$, while the boundary fractal dimension $D_{b}$ fluctuates narrowly around values between 1.0 and 1.2. Complementary to this behavior, the compactness parameter $\xi(t)$ exhibits only moderate fluctuations without sharp excursions.

In contrast, strain 106 exhibits pronounced, time-localized excursions in boundary-sensitive descriptors. Circularity decreases sharply from $C \sim 0.9$ to $C \sim 0.5$, followed by a gradual recovery over the interval $t \sim 49$--$100$~h, coincident with the morphological configurations shown in columns (III)--(IV) of Fig.~\ref{fig:morphology}(c). Over the same interval, $\xi(t)$ displays a sharp increase followed by a rapid decrease, while $D_{b}(t)$ exhibits a correlated enhancement. By comparison, the mass fractal dimension $D_{f}(t)$ evolves more slowly and remains near saturation throughout this period, indicating that substantial boundary reorganization can occur without a corresponding disruption of bulk space filling.

Strain 102 exhibits a qualitatively different temporal structure. Circularity increases rapidly from $C \sim 0.3$ to $C \sim 0.8$, accompanied by a strong decrease in compactness and a relaxation of the boundary fractal dimension from $D_{b} \sim 1.2$ toward $1.0$, reflecting rapid smoothing and filling-in of the initially irregular colony boundary. Following this initial transient, all boundary-sensitive descriptors enter a regime of stabilized evolution, marked by the vertical dashed line in Fig.~\ref{fig:shape}. Beyond this point, $C(t)$, $\xi(t)$, and $D_{b}(t)$ exhibit only weak temporal variation, while $D_{f}(t)$ continues to evolve slowly toward saturation. This behavior is consistent with the morphologies shown in columns (IV)--(V) of Fig.~\ref{fig:morphology}(d).

Taken together, these results reveal a separation of timescales between bulk space filling and boundary reorganization. While the mass fractal dimension evolves gradually across all strains, boundary-sensitive shape descriptors display either localized extrema associated with transient reorganization (strain 106) or rapid relaxation followed by stabilization (strain 102). In the next subsection, we show how these distinct modes of geometric evolution are reflected in the time-ordered area--perimeter scaling behavior.

\subsection{Time-ordered area--perimeter scaling of non-equilibrium growth morphology across strains}
\label{subsec:ap_scaling}

\begin{figure*}[htp!]
    \centering
    \includegraphics[width=0.9\textwidth]{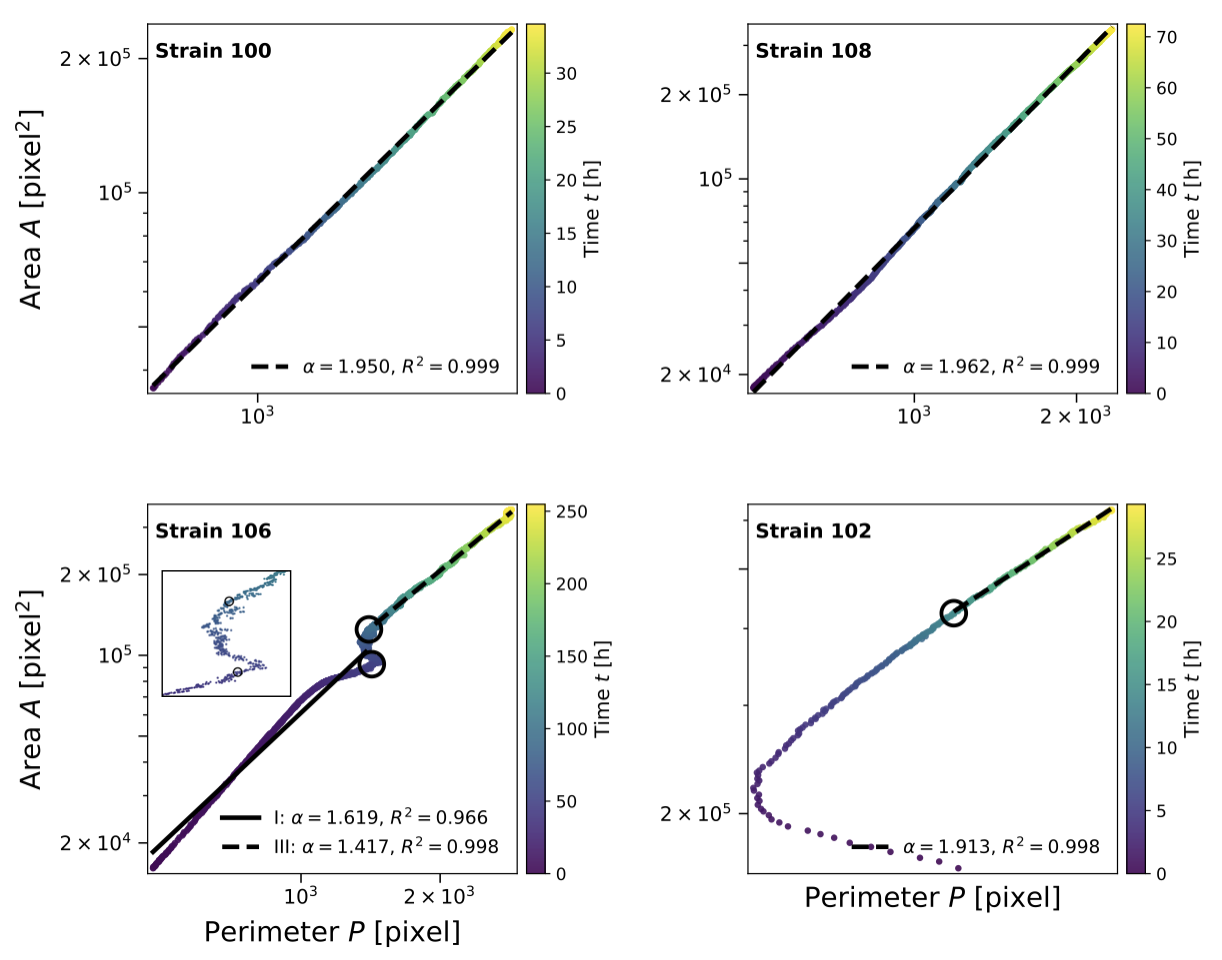}
    \caption{Time-ordered area--perimeter scaling of colony growth across strains. Log--log plots of colony area $A(t)$ versus perimeter $P(t)$, colored by time, for four bacterial strains (top row: strains 100 and 108; bottom row: strains 106 and 102). Strains 100 and 108 exhibit a robust single power-law relation $A \sim P^{\alpha}$ over the full growth trajectory, consistent with compact two-dimensional growth governed by a single geometric length scale. Strain 106 displays a clear crossover between two distinct scaling regimes (open circles), with the inset highlighting the intermediate region connecting these regimes. Strain 102 shows an early-time curved trajectory associated with rapid morphological reorganization, followed by the emergence of a stabilized power-law scaling regime at later times (open circle). For strain 106, the crossover region coincides with extrema in boundary-sensitive shape descriptors (Fig.~\ref{fig:shape}) and with the intermediate-time morphologies shown in Fig.~\ref{fig:morphology}(c).}

    \label{fig:ap-scaling}
\end{figure*}

We examine whether colony growth can be described by a single geometric length scale by analyzing the time-ordered relationship between area and perimeter over the full growth trajectories for all strains (Fig.~\ref{fig:ap-scaling}). By retaining temporal ordering, this representation captures the dynamical evolution of colony geometry rather than treating individual measurements as independent samples.

For strains 100 and 108 (top panels of Fig.~\ref{fig:ap-scaling}), the area--perimeter relation given by Eq.~\ref{eq:ap} gives an $\alpha \approx 2$ and $R^{2} > 0.99$ over the entire observation window. This behavior is consistent with compact two-dimensional growth, in which the perimeter provides the sole relevant geometric length scale governing area accumulation. Notably, despite pronounced differences in colony morphology and internal structure (Fig.~\ref{fig:morphology}a,b), both strains obey the same geometric constraint, demonstrating that visual heterogeneity does not necessarily imply a breakdown of compact growth. We observe this same trend for strain NCIB3610 as shown in Appendix~\ref{sec:strain-NC}.

In contrast, strain 106 does not admit a single-exponent description across its full trajectory (Fig.~\ref{fig:ap-scaling}, bottom left). Instead, the area--perimeter relation exhibits a clear crossover between two scaling regimes, characterized by $\alpha \approx 1.62$ ($R^{2} \approx 0.97$) at early times and $\alpha \approx 1.42$ ($R^{2} > 0.99$) at later times. The intermediate region connecting these regimes cannot be described by a single power law. Importantly, growth remains sustained throughout this interval: $A(t)$ increases monotonically, while $P(t)$ exhibits a transient excursion associated with boundary reorganization (Fig.~\ref{fig:AP-kinematics}). The crossover region, highlighted by open circles in Fig.~\ref{fig:ap-scaling}, coincides with extrema in boundary-sensitive shape descriptors (Fig.~\ref{fig:shape}) and corresponds to the intermediate-time morphologies shown in Fig.~\ref{fig:morphology}(c), spanning approximately $t \sim 49$--$100$~h.

Strain 102 exhibits a distinct two-stage behavior. At early times, the time-ordered $A$--$P$ trajectory follows a curved, non-scaling path associated with rapid morphological reorganization and boundary smoothing. Following this transient, the trajectory converges onto a well-defined power-law scaling regime, indicated by the open circle in Fig.~\ref{fig:ap-scaling}, with an exponent $\alpha \approx 1.9$ ($R^{2} \approx 0.998$). This transition coincides with the onset of stabilization in boundary-sensitive shape descriptors (Fig.~\ref{fig:shape}) and marks the emergence of effectively compact growth at later times, consistent with the morphologies shown in Fig.~\ref{fig:morphology}(d).

The exponent $\alpha \approx 2$ reflects growth controlled by a single characteristic geometric length scale: for a compact two-dimensional object, area scales with the square of the linear extent, while perimeter scales linearly. 

Taken together, Figs.~\ref{fig:AP-kinematics}, \ref{fig:shape}, and \ref{fig:ap-scaling} provide a unified dynamical geometric characterization of non-equilibrium colony growth. Sustained bulk expansion coexists with strain-dependent boundary reorganizations, which manifest as correlated extrema in boundary-sensitive shape descriptors and as transient departures from single-exponent area--perimeter scaling. The recovery or emergence of compact scaling at later times reflects the restoration of growth governed by a single dominant geometric length scale.

\section{Discussion and Outlook}
\label{sec:discussion}

In this work, we examined whether visually distinct bacterial colony morphologies necessarily reflect fundamentally different non-equilibrium growth behavior, or whether they can be understood within a shared geometric framework. To address this question, we adopted a geometry-first approach to colony growth, treating morphology as a dynamical observable reconstructed from time-resolved geometric measurements.

By combining measurements of area, perimeter, circularity, boundary as well as space-filling fractal dimension, and compactness, we show that colony growth is subject to strong geometric constraints. A central result is the identification of \emph{conditional geometric scaling}, in which a robust area--perimeter relation holds over extended time intervals but breaks down during transient episodes of geometric reorganization. Despite pronounced differences in colony appearance, three strains exhibit a robust area--perimeter relation consistent with compact two-dimensional growth governed by a single effective geometric length scale. In contrast, two other strains show a clear departure from this scaling, indicating a change in the geometric relationship between bulk accumulation and boundary evolution.

The observed breakdown of single-parameter area--perimeter scaling in strain 106 is consistent with growth geometries in which bulk accumulation and boundary advance are no longer controlled by a single characteristic length scale. Similar separations between interior filling and peripheral growth have been discussed in recent three-dimensional studies of biofilm expansion, where annular or ``napkin-ring''--like growth modes emerge~\cite{pokhrel}. While our analysis is restricted to two-dimensional projected geometry and does not resolve vertical structure, the present results suggest that such multi-scale growth modes can be detected directly from planar geometric observables, without assuming a specific three-dimensional growth model.

This deviation coincides with time-localized excursions in boundary-sensitive shape descriptors, linking the breakdown of scaling to transient geometric restructuring rather than noise or growth arrest. These results demonstrate that visual heterogeneity alone does not uniquely characterize non-equilibrium growth behavior, and that geometric observables provide a systematic means of distinguishing continuous growth from regimes of altered morphological evolution.


The observed behavior highlights a distinction between global geometric constraints and local morphological dynamics. Area--perimeter scaling probes whether growth is controlled by a single characteristic geometric length scale associated with the outer interface, while its breakdown signals the emergence of additional geometric structure that decouples perimeter growth from area accumulation. The more pronounced, time-localized variations in circularity and compactness arise from sensitivity to global shape anisotropy and transient boundary rearrangements. Together, these observations demonstrate that compact global growth can coexist with dynamically evolving, non-self-similar morphology, underscoring the need for multiple complementary geometric descriptors to characterize non-equilibrium growth. Importantly, these deviations do not arise from increased boundary roughness alone, but from a temporary decoupling between perimeter growth and bulk area accumulation, indicating the emergence of additional geometric degrees of freedom.

Here, we have focused on geometric observables that remain well defined without assuming a preferred growth direction, a parametrized interface, or a specific microscopic growth model. While alternative approaches based on local roughness exponents, interface velocities, or curvature statistics may yield complementary insights in idealized settings, our results show that key features of non-equilibrium growth—including compactness, scaling breakdown, and transient restructuring—can be identified directly from geometric measures alone.

Although the present analysis emphasizes the outer colony boundary, it is not restricted to purely interfacial information. The area $A(t)$ as well as space-filling fractal dimension $D_{f}$ capture effectively the integrated interior content of the colony and therefore reflect bulk growth and compaction in addition to boundary dynamics, highlighting the dynamic coupling between interior filling and interface organization at the colony scale. At the same time, the current framework treats the colony interior as spatially coarse-grained and does not resolve internal heterogeneity or subdomain organization.

Extending this geometric framework to incorporate internal morphology represents a natural next step. Identifying internal growth fronts, density gradients, or morphological domains would enable investigation of how interior organization couples to boundary-driven growth and whether additional constraints or transitions emerge at the level of internal structure.


\section*{Acknowledgments}
SS received financial support from Brandeis University. BH acknowledges the Vermont Advanced Computing Center (VACC) at the University of Vermont for providing computational resources that have contributed to the research results reported within this paper.

\appendix

\section{Robustness and parameter dependence of fractal-dimension estimates}
\label{sec:appendix_fractal}

We assess the robustness of the fractal-dimension measurements used in the main text by examining the dependence of the mass fractal dimension $D_f$ and boundary fractal dimension $D_b$ on analysis parameters for a representative strain. In particular, we vary the box-size range and the thickness of the binary mask used in the box-counting procedure, which controls the effective discretization of the colony interior and boundary.

\begin{figure}[htp!]
    \centering
    \includegraphics[width=0.95\columnwidth]{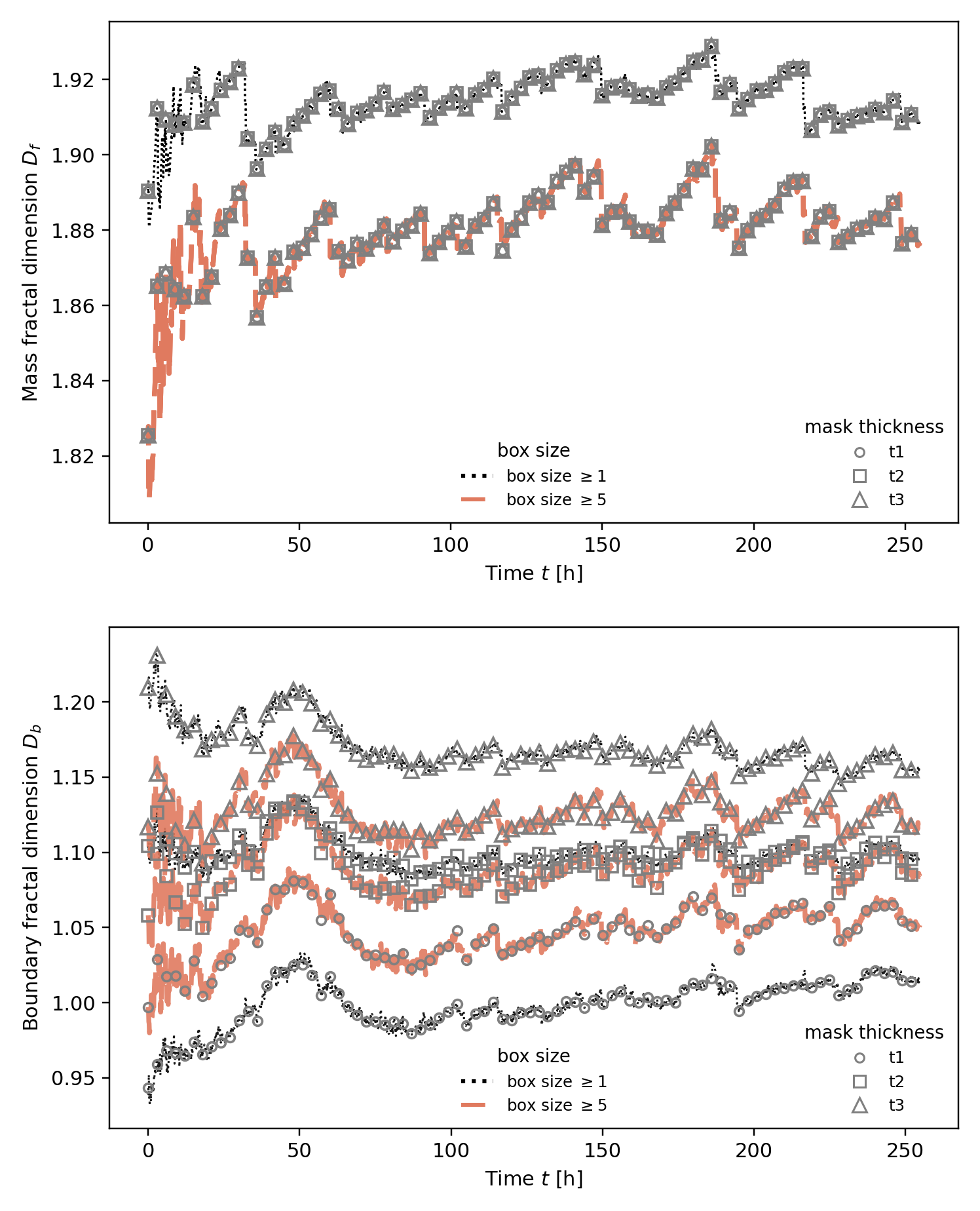}
    \caption{Dependence of fractal-dimension estimates on analysis parameters for Strain 106. 
    \textit{Left panel:} Mass fractal dimension $D_f$ as a function of time for three effective thickness values and two box-size ranges used in the box-counting analysis. 
    \textit{Right panel:} Boundary fractal dimension $D_b$ under the same conditions. 
    Across all parameter choices, $D_f$ rapidly approaches a plateau near values characteristic of compact two-dimensional growth $\sim2$, while $D_b$ evolves smoothly within a narrow range reflecting boundary roughness. 
    Although the absolute values depend modestly on box size and thickness, the qualitative temporal trends are robust.}
    \label{fig:fractal_appendix}
\end{figure}

Figure~\ref{fig:fractal_appendix} shows the evolution of $D_f$ and $D_b$ as a function of time for three effective thickness values (thickness=1,2,3) and two box-size ranges (box-size$\ge1$, box-size$\ge5$). The mass fractal dimension $D_{f}$ shows no dependence on thickness and only a weak dependence on box size, rapidly approaching a plateau at values characteristic of compact two-dimensional growth $\sim2$. Across all parameter choices, $D_f$ varies smoothly and remains bounded, indicating that internal space filling evolves gradually and is insensitive to transient morphological rearrangements.

By contrast, the boundary fractal dimension $D_b$ shows a modest dependence on box size and thickness. As expected for a boundary-sensitive measure, for Strain 106 $D_b$ increases gradually over time till 50 h and then decreases and remains confined to a narrow range ($D_b \sim 1.0$--$1.3$), reflecting multiscale roughness of the colony edge. While localized enhancements in $D_b$ can coincide with periods of increased boundary corrugation discussed in the main text, the overall evolution of $D_b$ remains smooth and does not exhibit sharp discontinuities across parameter choices. Taken together, these results demonstrate that fractal-dimension estimates are robust to reasonable variations in analysis parameters.

\section{Geometry \& dynamical morphology of strain NCIB3610}
\label{sec:strain-NC}
\begin{figure*}[htp!]
    \centering
    \includegraphics[width=0.98\linewidth]{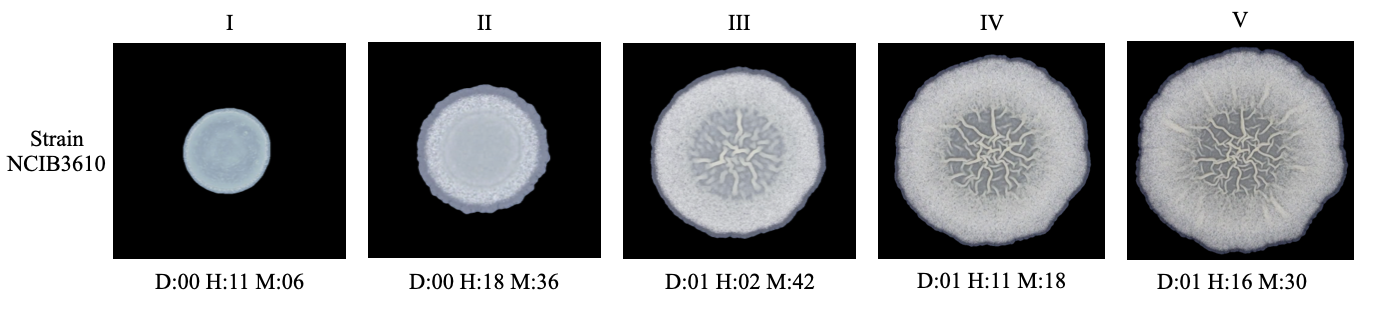}
    \caption{Time-resolved snapshots of colony growth morphology for the \emph{Bacillus subtilis} strain NCIB3610.
Five representative time points are shown, labeled (I)--(V), progressing from early to late stages of growth.
The displayed colony footprint corresponds to the same extracted geometry used for subsequent quantitative analysis.
These snapshots illustrate the temporal evolution of colony shape, boundary organization, and internal patterning across the growth process.
}
    \label{fig:morph-nc}
\end{figure*}

\begin{figure*}[htp!]
    \centering
    \includegraphics[width=0.95\linewidth]{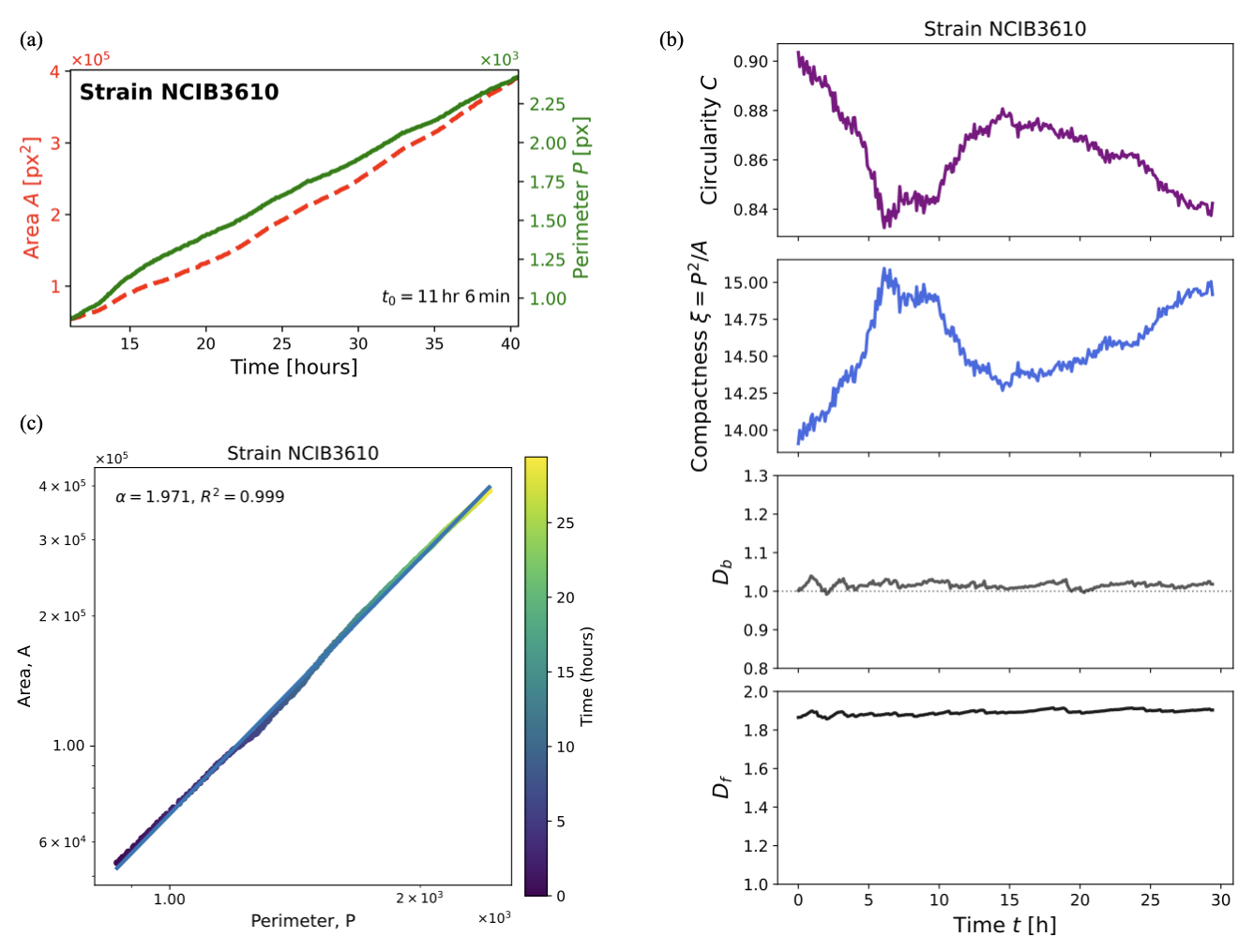}
    \caption{(a) Time evolution of colony area $A(t)$ (red, dashed) and perimeter $P(t)$ (green, solid) as extracted from time-resolved microscopy images. Both quantities increase smoothly over time, indicating sustained growth. (b) Temporal evolution of shape descriptors: circularity $C(t)$ (first row, purple), compactness $\xi(t)=P^2/A(t)$ (second row, blue), boundary fractal dimension $D_{b}$ (third row, gray) and space-filling fractal dimension $D_{f}$ (fourth row, black) showing gradual boundary roughening accompanied by modest, time-dependent variations in global shape measures. (c) Time-ordered area–perimeter relation shown on log–log axes, colored by time, exhibiting a robust power-law scaling $A \sim P^{\alpha}$ with $\alpha \approx 1.97, R^{2}>0.999$ over the full observation window, consistent with compact two-dimensional growth governed by a single effective geometric length scale. Together, these results demonstrate that strain NCIB3610 exhibits compact growth with well-defined geometric scaling and constrained evolution in morphology space, consistent with the behavior observed for strains 100 and 108 in the main text.}
\label{fig:NC-params}
\end{figure*}

In this section, we provide the time-resolved morphological images (see Fig.~\ref{fig:morph-nc}) along with the time evolution of area-perimeter, effective boundary and mass/space-filling fractal dimension, circularity and compactness for strain NCIB3610 (see Fig.~\ref{fig:NC-params}). We also show the validity of the area-perimeter scaling with $\alpha=1.971, R^{2}>0.999$ (see Fig.~\ref{fig:NC-params} for details.) We observe that while strain NCIB3610 looks different in the time-resolved morphological evolution than strains 100 and 108, it obeys the area-perimeter scaling for a radial compact growth equivalent to the other two strains. Calculations for effective fractal dimensions are performed with thickness=1 and box-size $\ge 5$. 

\clearpage
\bibliography{refs_corrected}
\end{document}